\documentclass[10pt, journal, a4paper, onecolumn]{IEEEtran}
\pdfoutput=1

\usepackage{cite}

\usepackage{xcolor,colortbl}

\definecolor{Gray}{gray}{0.85}
\newcolumntype{g}{>{\columncolor{Gray}}c}

\ifCLASSINFOpdf
   \usepackage[pdftex]{graphicx}
\else
  \usepackage[dvips]{graphicx}
\fi
\makeatletter
\let\MYcaption\@makecaption
\makeatother

\usepackage[font=footnotesize]{subcaption}

\makeatletter
\let\@makecaption\MYcaption
\makeatother

\usepackage{url}

\usepackage{todonotes}

\newcommand{\cits}[0]{\mbox{C-ITS}}

\begin{document}
\title{Cellular Network Multi-Access Measurements \\on the Roads of V\"{a}rmland, Sweden}

\author{%
\IEEEauthorblockN{Fehmi Ben Abdesslem, Henrik Abrahamsson, and Bengt Ahlgren\\}
\IEEEauthorblockA{RISE SICS\\
fehmi.ben.abdesslem@ri.se, 
henrik.abrahamsson@ri.se, 
bengt.ahlgren@ri.se
}}


\maketitle

%
%
\newenvironment{note}{\begin{quote}\sffamily}{\end{quote}}
\def\comment#1{\leavevmode\marginpar{\raggedright\scriptsize\sffamily #1}\ignorespaces}
\newcount\hour
\newcount\minute
\hour=\time\relax
\divide\hour by 60\relax
\minute=\hour\relax
\multiply\minute by -60\relax
\advance\minute by \time\relax
\def\timeprint#1{\ifnum #1<10 0\the#1\else \the#1\fi}

\begin{abstract}

Cooperative Intelligent Transport Systems (C-ITS) make road traffic safer and more efficient, but require the mobile networks to handle time-critical applications. While some applications may need new dedicated communications technologies such as IEEE 802.11p or 5G, other applications can use current cellular networks. This study evaluates the performance that connected vehicles can expect from existing networks, and estimates the potential gain of multi-access by simultaneously transmitting over several operators. We upload time-critical warning messages from buses in Sweden, and characterise transaction times and network availability. We conduct the experiments with different protocols: UDP, TCP, and HTTPS. Our results show that when using UDP, the median transaction time for sending a typical warning message is $135$~ms. We also show that multi-access can bring this value down to $73$~ms. For time-critical applications requiring transaction times under $200$~ms, multi-access can increase availability of the network from to $57.4$\% to $92.0$\%.


\end{abstract}

\IEEEpeerreviewmaketitle

\section{Introduction}
\label{sec:introduction}
Connected cars and cooperative intelligent transport systems (\cits) can make road traffic safer and more efficient. When vehicles and the road infrastructure collect and share information about the planned routes and the surrounding environment, it opens up for many new services and applications. Drivers can for example get early warnings about road hazards ahead, and advanced driver assistance systems can automate braking to avoid collisions. \cits\ is also expected to enhance and support autonomous driving. 

Many future \cits\ applications may require new communications systems such as 5G or a whole new infrastructure dedicated to vehicular networks based on the new IEEE 802.11p standard. However various applications can already leverage the current cellular network infrastructure and their deployment should not be delayed until the new systems are deployed. Even then, remote areas might take longer to be covered and will most likely rely on existing cellular networks. Relying on existing networks for \cits\ has been explicitly recommended by public authorities. For example, the \cits\ Platform\footnote{\url{https://ec.europa.eu/transport/themes/its/c-its_en}} set up in 2014 by the European Commission advocates in its initial report (2016)\cite{EU:cits16} the use of the existing cellular communications infrastructure in order to foster uptake of \cits\ services, before the future deployment of short-range communications in the 5.9 GHz band described in standards such as ETSI~\mbox{ITS-G5}. This same report recognises the many uncertainties related to using existing cellular networks for \cits\ services, including coping with latency-critical services. This work contributes to assess and reduce those uncertainties. In its second report (2017)\cite{EU:cits17}, the \cits\ Platform recommends following a \emph{hybrid communication approach} where cellular networks are not only a temporary solution but also a complementary infrastructure to be used along other future technologies, hence confirming the long-term relevance of cellular networks for \cits\ services.  


This work focusses on a simple scenario, common to many \cits\ applications, where a vehicle sends data to a server and receives a reply. For instance, location, destination, speed, or surrounding events (potentially hazardous) are sent to a server, which replies in return with a new route or warnings based on data collected from other sources, such as other vehicles. The time constraint on the transaction varies from tens of milliseconds to seconds depending on the application~\cite{Karagiannis:cst11}. 


To improve the design of \cits\ time-critical applications and better understand what can be expected from the underlying cellular networks, the performance of such networks needs to be measured. To do so, we ran experiments on the road where we measured the availability of the network and the transaction time when messages are uploaded and acknowledged by a server. Transactions were performed from a vehicle in motion to a server via different cellular network operators. To the best of our knowledge, this is the first study measuring the performance of uploading data from vehicles in uncontrolled settings and for different operators.

The main contribution of this work is two-fold. Firstly, we experimentally characterise the performance of uploading data on the road by sharing our insights on the data collected from our experiments. Such characterisation can be used to feed simulators with realistic distributions of performance that can be offered by the network. Designers of applications and services can also benefit from our analysis to better understand what they can expect from their application when deployed in the wild. Secondly, we evaluate the potential gain of using a \emph{multi-access} approach, which consists in sending messages on several mobile networks. Examples of our main results are:
\begin {itemize}
\item When using UDP, the median transaction time for sending a typical warning message is $135$~ms.
\item Multi-access can bring this value down to $73$~ms.
\item For time-critical applications requiring transaction times under $200$~ms, multi-access can increase availability from to $57.4$\% to $92.0$\%.
\end{itemize}

The remainder of this technical report is organised as follows. First, Section~\ref{sec:related} outlines related works. Section~\ref{sec:measurementsetup} then describes the measurement platform and the experiment design. After presenting the results in Section~\ref{sec:results}, a discussion on their possible implications and on ongoing work is provided in Section~\ref{sec:discussion}. The paper is concluded in Section~\ref{sec:conclusions}.

\section{Related Work}
\label{sec:related}
The development of \cits\ is moving fast. Various projects have set roadmaps and promoted \cits\ development~\cite{codecs,car2car}, while other projects already started deploying and testing solutions on the road~\cite{korridor,nordicway,ukcite}. 

Karagiannis \textit{et al.} \cite{Karagiannis:cst11} survey the main use-cases and applications expected to be deployed with \cits.
The authors recognise that those applications require the underlying communication network to guarantee strict time constraints, sometimes under $100$~ms.  Lu \textit{et al.}\cite{Lu:iotj14} survey the different solutions that have been proposed to ensure the wireless communication between vehicles (V2V) and the road infrastructure (V2I).

While IEEE~802.11p is the de facto standard expected to support \cits\ applications, it requires new road-side equipment to be installed to communicate in the 5.9~Ghz band. The research community recognises the need to rely on current cellular networks to bootstrap the uptake of \cits\ applications instead of waiting for a full deployment of IEEE~802.11p networks~\cite{Araniti:commag13}. In parallel, measuring different aspects of the cellular networks has been the focus of a large set of studies. Albadejo \textit{et al.}\cite{Albaladejo:pimrc16} measure the downlink bandwidth and RTT (Round Trip Time) at different fixed locations in Dublin. Huang \textit{et al.} measure the maximum downlink and uplink bandwidth from 20 smartphone users over five months~\cite{Huang:Mobisys12}. Sommers \textit{et al.}\cite{Sommers:imc12} compare the performance of cellular networks and WiFi networks from a crowd-sourced dataset collected from a speed test application for mobile phones. Xu \textit{et al.}\cite{Xu:mobisys13} analyse cellular network traces from three different locations and show the predictability of network conditions. Finally, other works\cite{Huang:sigcomm13,Garcia:wiopt14,Garcia:eucnc15} more generally focus on characterising TCP in cellular networks.

Improving latency has been addressed by several works\cite{briscoe:survey14}, particularly in cellular networks\cite{Jiang:imc12}. To this end, redundancy is one of the popular promising solutions, as proposed by Vulimiri \textit{et al.}\cite{Vulimiri:hotnets12}. Another solution is to use multi-path by combining cellular and WLAN interfaces\cite{Yedugundla:comnet16,Chen:imc13,Ferlin:globecom14}.

Recently, studies have focused on the connectivity pattern of vehicles\cite{Andrade:imc17}. Khatouni \textit{et al.}~\cite{Khatouni:itc17} measured the downlink performance and RTT for different cellular networks, including in buses, using the MONROE platform~\cite{Alay:wowmom16,Alay:mobicom17}. 

Our work is at the cross roads of the afore-mentioned related works, as it addresses the measurement of cellular network performance and aims at assessing the potential for reducing the transaction time. We target applications for vehicles but our work differs from the general C-ITS research and pilot studies in that: (i) we focus on the uplink communication performance and not in the downlink; (ii) we evaluate performance in terms of transaction time and availability, and not in terms of bandwidth or single packet RTT; (iii) we study the potential gain of multi-access by transmitting over different cellular network operators; (iv) we collect and analyse data from nodes on the road, transmitting from within vehicles in motion on existing commercial networks.

\section{Scenario and Experiments}
\label{sec:measurementsetup}
\begin{figure}[h]
\centering
\includegraphics[width=.5\hsize]{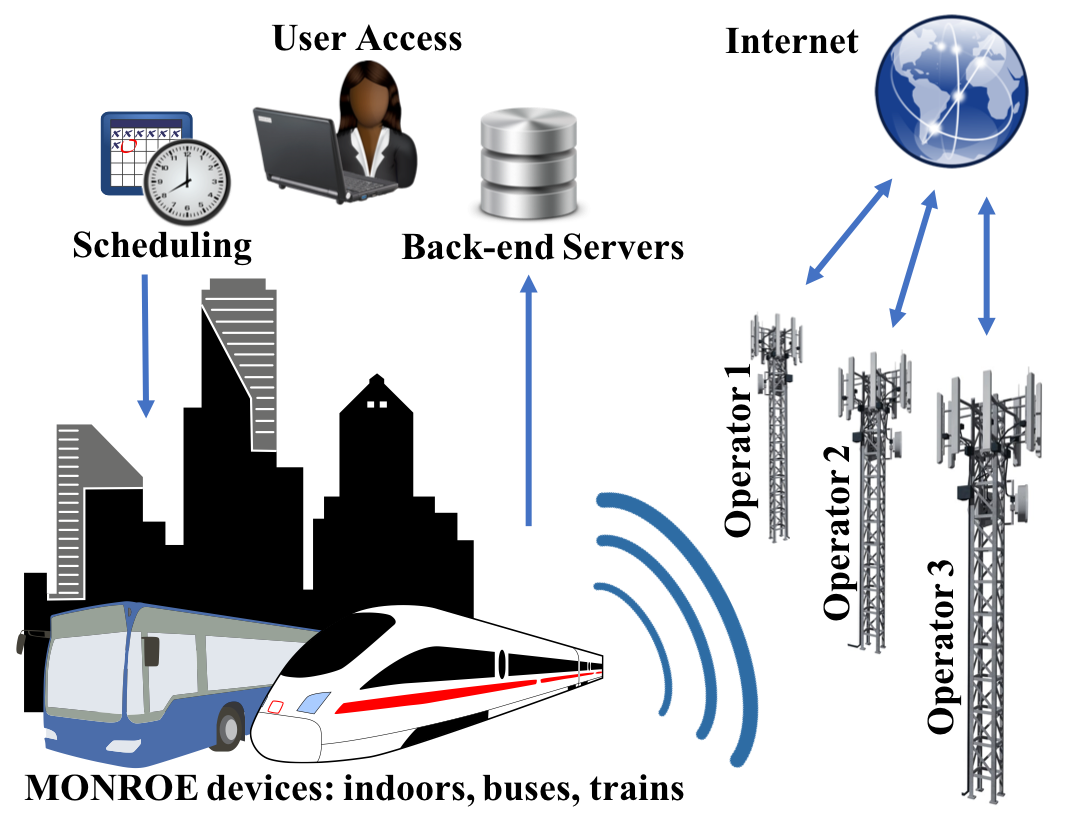}
\caption{The MONROE mobile broadband measurement platform.}
\label{fig:monroe_overview}
\end{figure}

\begin{figure*}[h]
    \includegraphics[width=\textwidth]{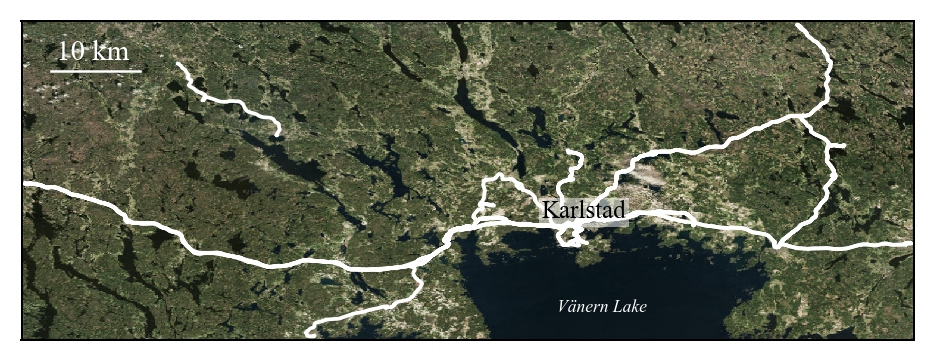}
    \caption{Satellite map showing the routes covered by buses in our experiments, in the Swedish region of V\"{a}rmland.}
    \label{fig:map}
\end{figure*}

In this section, we define the scenario envisioned in our work, before
describing in detail the measurement platform and the experiments.

\subsection{Scenario}

Our scenario is based on the concept of Floating Car Data
(FCD~\cite{Pfoser2008,oberstein1997collection,schafer2002traffic}), where data generated by vehicles are collected, typically telemetric data such as speed,
direction and position of the vehicle. FCD is used to assess the overall
traffic conditions and has been implemented for more than two decades. Popular
services nowadays such as Google Maps Navigation and Waze still use such
method, with smartphones in vehicles~\cite{jeske2013floating}. 
While uploading data, road users also receive traffic flow information, allowing them to avoid
congested areas and optimise their route~\cite{de2008traffic}. By
crowd-sourcing the data collection, FCD also reduces the need for deploying
traffic monitoring infrastructure. Hence, as more and more vehicles get
connected and \cits\ services are deployed, FCD is gaining in popularity. 
Extended FCD (EFCD, or xFCD) relies on the FCD concept but adds to collected
data any other data generated by the different embedded equipment in the
vehicles~\cite{huber1999extended}. New C-ITS services could rely on EFCD. For example, hazard warning signals could be sent to the network and warn other road users, even when they are not in line of sight. 
The European Telecommunications Standards Institute (ETSI) provides technical specifications
on how to send EFCD between different ITS stations (in-car stations, roadside
stations, and central remote stations) via ITS-G5 or cellular
networks~\cite{EU:TS102}. For ITS-G5 networks, the ETSI proposes EFCD to first
be uploaded to roadside stations in range, which aggregate and send it to
central ITS stations via the internet in the European standard format, Datex
II.\footnote{http://www.datex2.eu/}  

We envision that until ITS-G5 or similar technologies are deployed, EFCD will
fully or partially rely on cellular networks to be uploaded directly from
vehicles to central servers, without any roadside station aggregating and
relaying the information. Very often, the transaction would require an answer
from the server: either to acknowledge reception of the transferred data, or to
provide corresponding information to the car such as an updated route.

In this work, we consider a scenario where a vehicle on the road uploads EFCD
in Datex~II format over available cellular networks to a server on the Internet.
The server then replies with a short message. We are interested in measuring
the transaction time in such scenario for different protocols, operators, and
EFCD sizes.

\subsection{Measurement platform}

For our measurements, we use the MONROE platform (Measuring Mobile Broadband
Networks in Europe ~\cite{Alay:mobicom17,Alay:wowmom16}), a distributed
platform for measuring, monitoring and assessing the performance of mobile
broadband services. MONROE consists of a testbed with hundreds of measurement
devices deployed over Europe. Many of the measurement devices are deployed on
buses, trucks and trains in motion. The platform is open to external
researchers and provide Experiment-as-a-Service. The measurement devices have
access to multiple operators which gives the possibility to compare operators
and to do multi-access experiments. Fig.~\ref{fig:monroe_overview} outlines
the high-level system design of the platform: users deploy experiments with a
scheduling tool to the measurement devices, each of which are typically equipped with
three LTE modems to transmit data over different commercial mobile broadband
(MBB) networks. The measurement devices then upload results of
the experiments into back-end servers, accessible by users for analysis.

\subsection{Experiment design}
\label{subsec:exp}
In this work we use the MONROE platform to study what communication performance connected vehicles can expect in current 3G/4G networks. We measure transaction performance using UDP, TCP and HTTPS from buses in motion.

\begin{figure}
\centering
\begin{minipage}{.45\linewidth}
    \begin{subfigure}{\textwidth}
        \includegraphics[width=\textwidth]{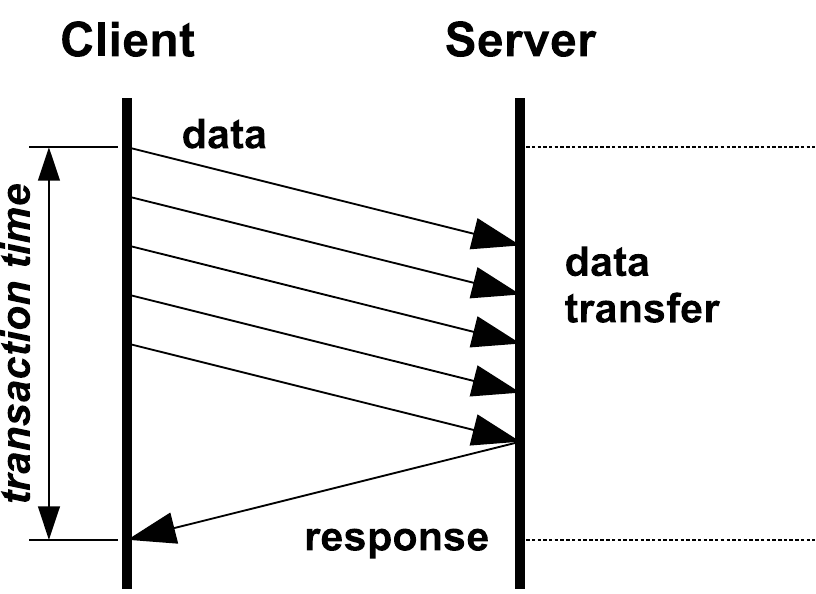}
        \caption{UDP}
    \end{subfigure}\\[2ex]
    \begin{subfigure}{\textwidth}
        \includegraphics[width=\textwidth]{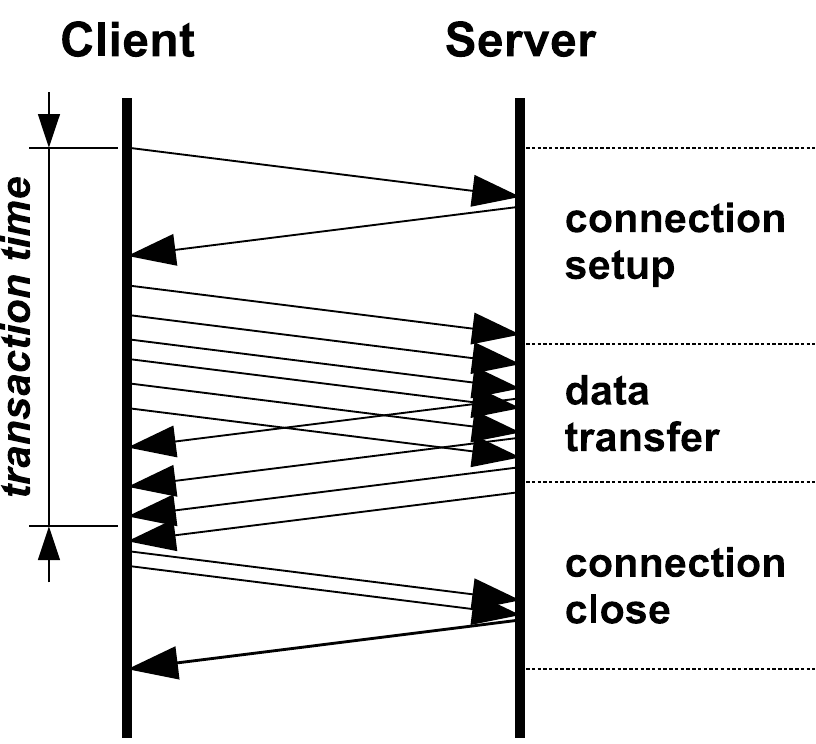}
        \caption{TCP}
    \end{subfigure}
\end{minipage}%
\hspace{2em}%
\begin{minipage}{.45\linewidth}
\begin{subfigure}{\textwidth}
        \includegraphics[width=\textwidth]{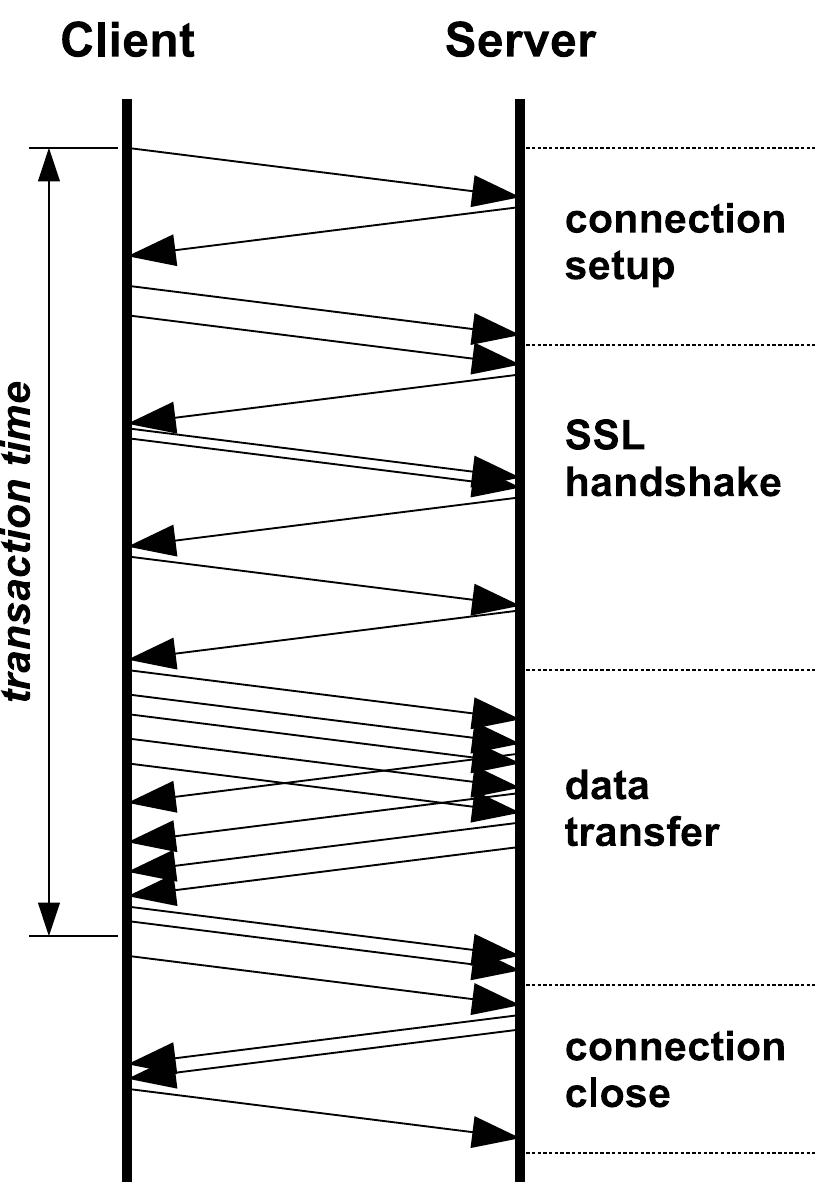}
      \caption{HTTPS}
\end{subfigure}
\end{minipage}

    \caption{Time sequence diagrams illustrating a transaction for each protocol.}
    \label{fig:timesequence}
\end{figure}

Using the MONROE platform we perform multi-access experiments with
duplicate transactions in parallel over the three available operators. The
objective is to study if, and to what extent, multi-access can decrease transaction time
and increase availability.

To upload a message with a realistic size, we design an experiment in which we emulate a time-critical \cits\ application that gets a sensor
reading at time $t$ and attempts to upload a warning message to the server
as fast as possible. We format the warning message with the Datex II XML standard, embedding a \textit{situationRecord} element of type \textit{VehicleObstruction} and containing identifiers, timestamps, and location coordinates. To authenticate the message, the XML message includes a signature with an X509 certificate. The total size of the message to upload adds up to $5.6$ KB. We duplicate this warning message and upload it over the three available
cellular network operators in parallel. 
The server knows what total size of data to expect, and hence replies
to the client as soon as all data is received. The reply is an acknowledgement message containing the number of bytes successfully received by the server. On the server side, we set a time-out value of five~seconds, after which when no more packets are received despite expecting more data, the server sends back the reply prematurely with the number of bytes received. On the client side, after sending all packets of the message, we set a time-out value of six~seconds after which when no answer is received from the server, we consider the message to be lost and the transaction unsuccessful. Those time-out values prevent both client and server from stalling while waiting for an answer.

We define a \emph{transaction} as the sequence of events starting with the transmission of the message being uploaded to the server, and ending either: (i) with the reception by the client of the reply sent by the server; or (ii) with the expiration of the time-out set by the client when no reply is received. We define a transaction as a \emph{successful transaction} if, and only if: (i) no time-out has expired throughout the transaction, neither on the client side, nor on the server side; \textit{and} (ii) the answer from the server as received by the client contains the same value as the amount of data sent by the client.

We chose to perform the transactions using three different protocols that were selected for their varied network characteristics and complexity: UDP for its light and connectionless model, TCP for its reliability mechanisms, and HTTPS for its SSL/TLS encryption and authentication features. Trade-off between performance and features depends on the application. Our goal is not to compare protocols between each other, but to show their performance.

The User Datagram Protocol (UDP) is a transport layer protocol using a connectionless communication model. 
Since it includes no mechanism to palliate the loss of packets, we envision its use for time-critical \cits\ applications where messages are not important but should still be received quickly to stay relevant. 
The Transmission Control Protocol (TCP) is also a transport layer protocol, but provides a reliable, ordered, and error-checked delivery of the messages. This protocol is more likely to be used by time-critical \cits\ applications for important messages, such as safety-related warnings, that should be received quickly and remain relevant even when delayed. 
Finally, HTTPS is a popular application layer protocol using TCP, that, despite its complexity, is likely to be used by \cits\ applications as it is largely supported and includes SSL/TLS encryption and authentication mechanisms.

The measurements for each of the three protocols were implemented by a couple of custom Python programs, respectively for the client and server. The Python program on the client side for HTTPS instruments a popular command line tool called \texttt{curl}\footnote{\url{https://en.wikipedia.org/wiki/CURL}} that sends the message using the HTTP POST method and embeds a TLS certificate. The Python program on the server side for HTTPS is behind an instance of the \texttt{nginx}\footnote{\url{https://en.wikipedia.org/wiki/Nginx}} web server, which handles the TLS client authentication. For all protocols, each measured transaction is logged on the client side with the transaction time and a number of metadata values from the MONROE platform such as location coordinates and signal quality. In addition, packet traces are captured both on the client and server sides to allow detailed analysis on the packet level.

We divide the experiments into \textit{experiment runs}. One experiment run lasts one hour and is divided into $120$ \textit{rounds}. A round starts at time $t$ and lasts $30$ seconds. At time $t + 10$ seconds, a UDP transaction is scheduled to be executed with each of the three available operators in parallel. A TCP transaction is scheduled at $t + 20$ seconds, and an HTTPS transaction at $t + 30$ seconds, hence ending the round with a total of nine transactions. Over the one hour of an experiment run, the $120$ rounds consist of $1,080$ transactions in total. 

We have run the experiments on mobile MONROE devices on buses operating in the city of Karlstad, Sweden, and surroundings. The geographical area covers urban, suburban, and rural zones, as shown in Fig.~\ref{fig:map}. All experiments have been executed while the buses where in motion. To increase the probability of the buses being in motion, we ran experiments on business days from the morning to early evening. The server is stationary and located in Stockholm, Sweden.

\subsection{Metrics}

We focus on two main metrics to evaluate the performance of each protocol: (i) the \emph{transaction time}, and (ii) the \emph{availability}. We define both metrics hereafter.

\emph{Transaction time}. Fig.~\ref{fig:timesequence} illustrates a typical transaction without any packet loss between the client in a bus and the server, for a $5.6$ KB message and for each protocol. We measure the transaction time from the client. For UDP the transaction time covers the complete packet exchange, including data upload and server response, while for TCP and
HTTPS, we exclude the connection closure. Intuitively, UDP is more likely to perform the transaction faster than TCP, which in turn will be faster than HTTPS, because of the increased complexity. For UDP, the transaction only requires one round trip between the
client and the server. For TCP, there are at least two round trips,
one for the handshake setting up the connection, and one for the data transfer. In
the illustrated case with a $5.6$ KB message, all data can be
transferred in the initial TCP window, before waiting for the acknowledgement packets. Finally, for HTTPS, the transaction typically includes five round trips, with three round trips for the TLS handshake only, before the data transfer. 

\emph{Availability}. We define the availability, or success rate, as the proportion of successful transactions out of the total number of attempted transactions. For example, an availability rate of $100$ \% means that all transactions were successful. We calculate this metric for each protocol, and possibly with a time constraint: if the transaction time exceeds a time limit required by a hypothetical \cits\ application, we consider the transaction as unsuccessful.

\section{Results}
\label{sec:results}

This section presents the results obtained from our experimental setup as described in the previous section. We performed $60$~experiment runs as described in Section~\ref{subsec:exp}. During the experiments, the buses on the road have attempted to upload $21,600$~messages of $5.6$~KB for each protocol and over $60$~hours. Since we focus on evaluating the performance of transmitting via different operators at the same time to study the benefit of multi-access, we only keep rounds of transactions where transactions started within $10$~ms from each other. For each transaction, we collected the measured transaction time, the status of that transaction, and related meta-data: GPS location coordinates, name of the operator, and information about the state of the network such as the signal strength. 

\begin{figure}[!h]
    \begin{subfigure}[t]{.45\linewidth}
        \includegraphics[width=\linewidth]{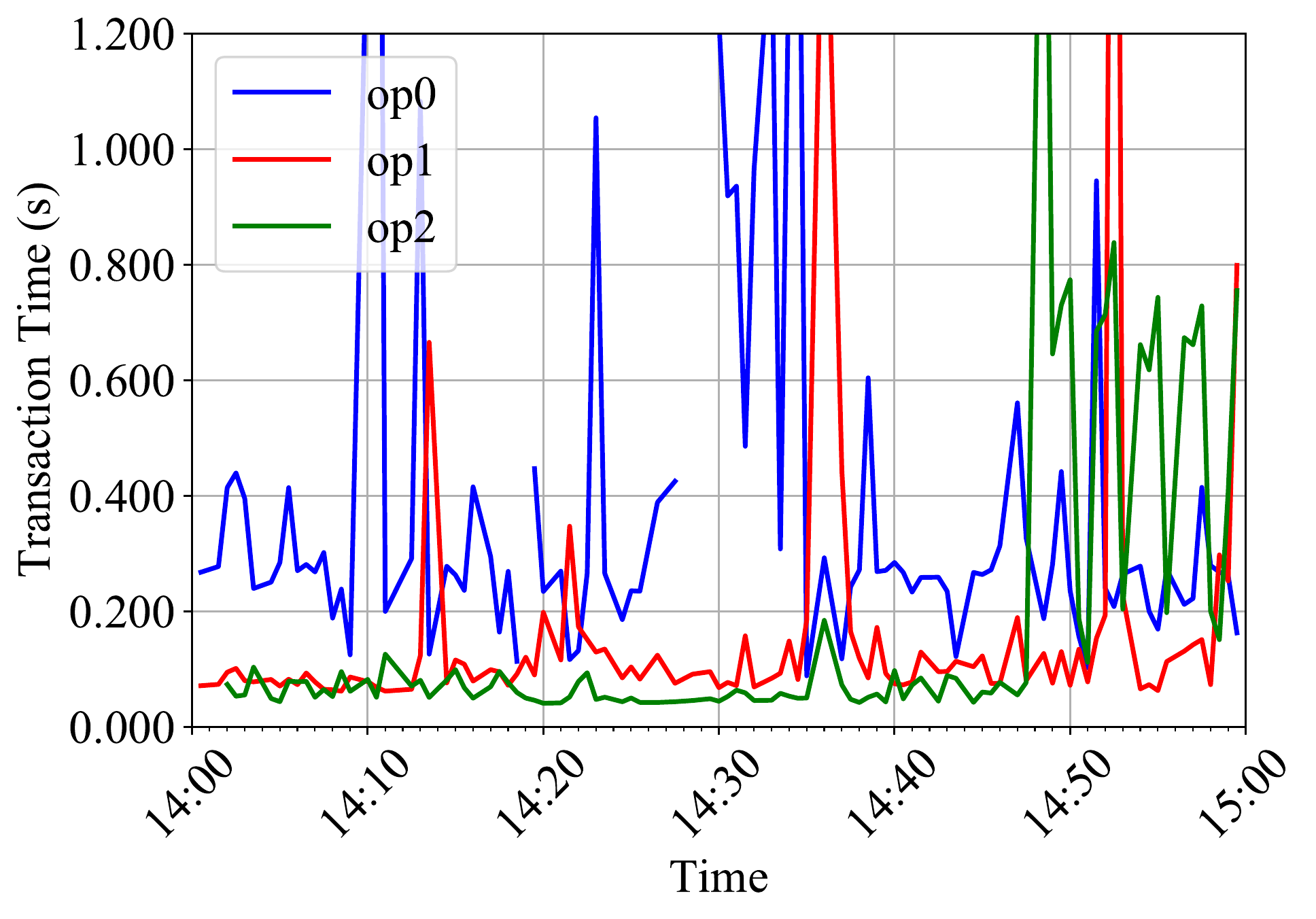}
    \end{subfigure}
    \begin{subfigure}[t]{.45\linewidth}
        \includegraphics[width=\linewidth]{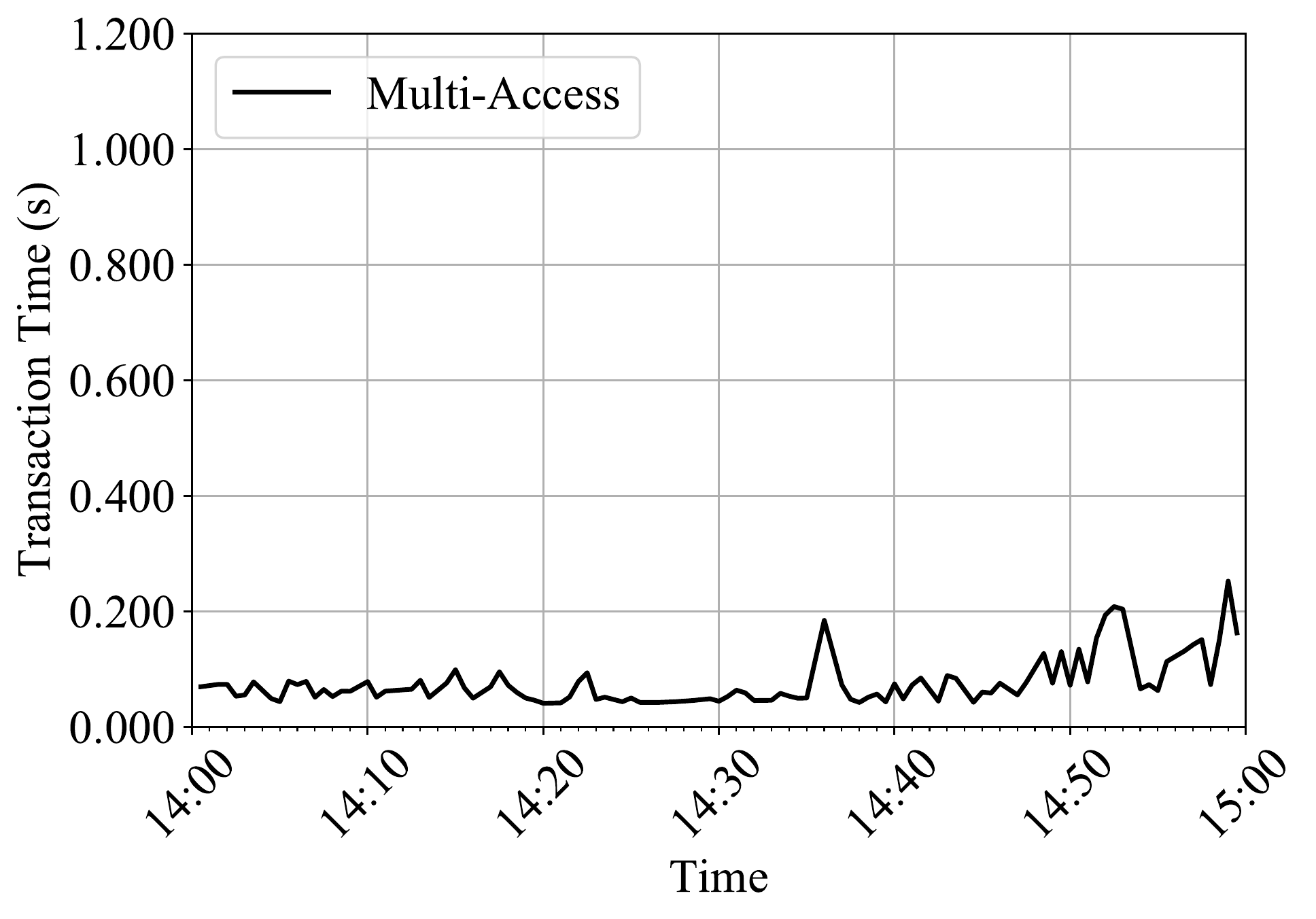}
    \end{subfigure}
    \caption{Typical example of an experiment run with UDP, illustrating how multi-access reduces the overall transaction time while increasing availability.}
    \label{fig:example}
\end{figure}
\begin{figure}[!h]
    \begin{subfigure}[t]{.45\linewidth}
        \includegraphics[width=\linewidth]{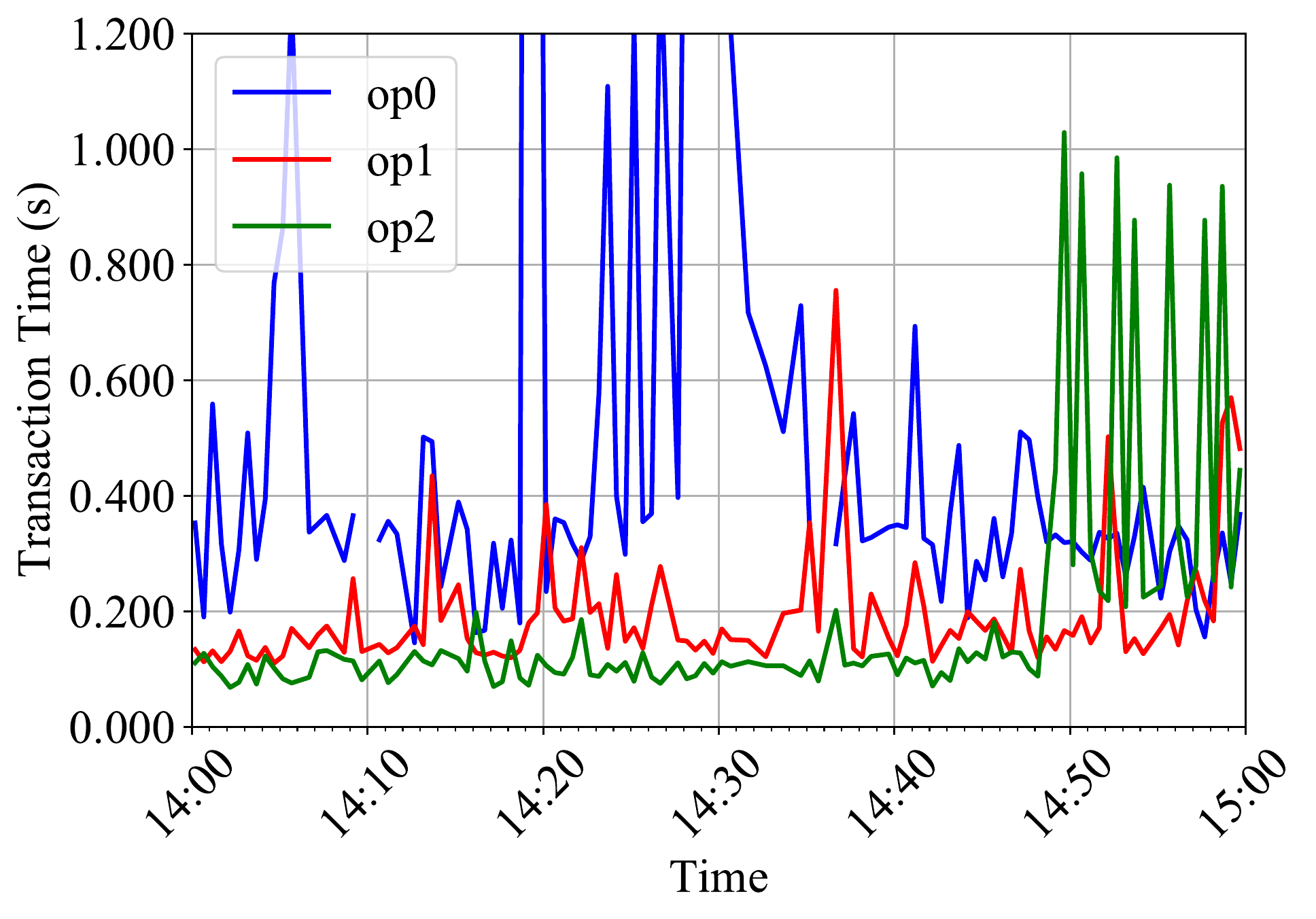}
    \end{subfigure}
    \begin{subfigure}[t]{.45\linewidth}
        \includegraphics[width=\linewidth]{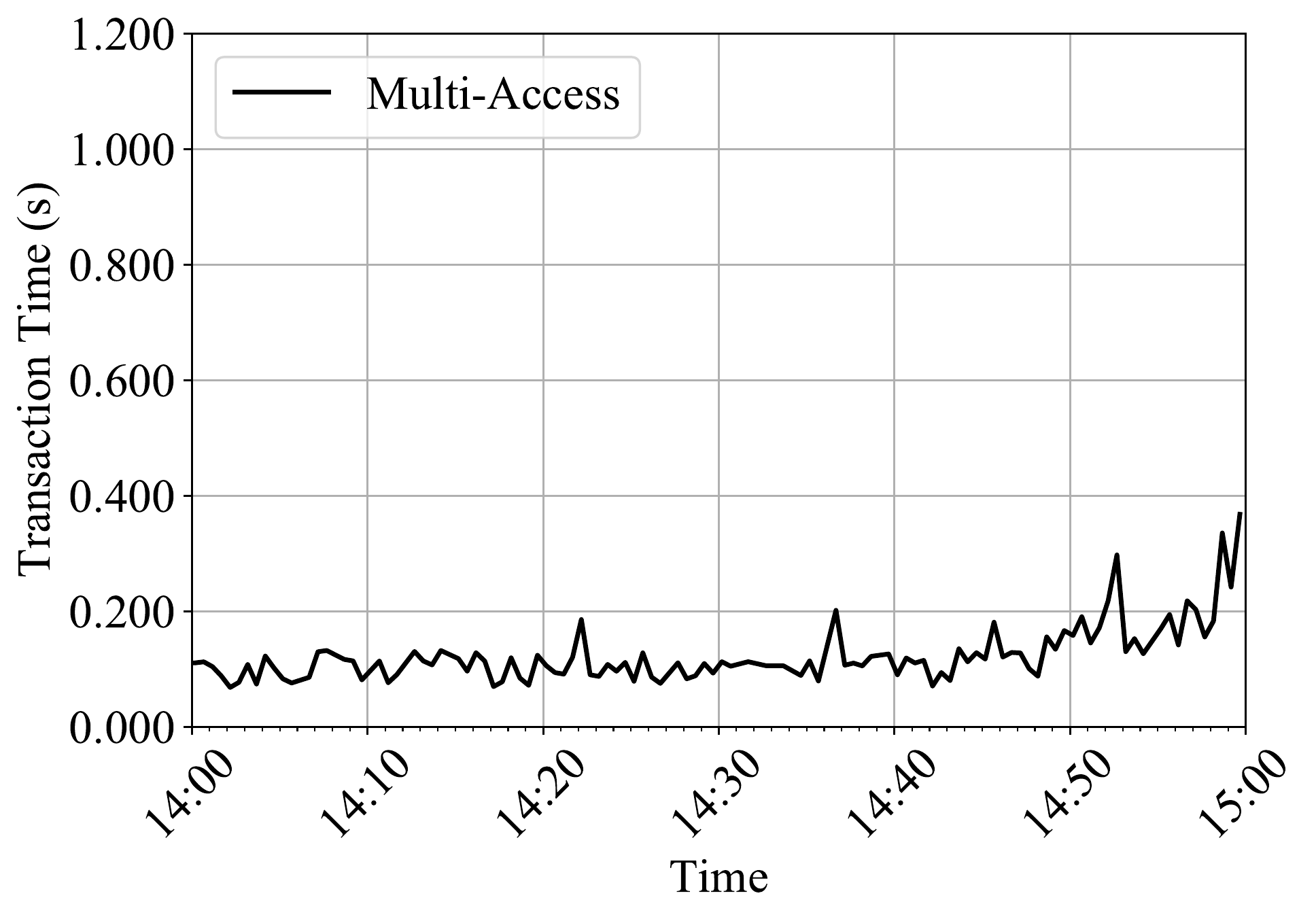}
    \end{subfigure}
    \caption{Typical example of an experiment run with TCP.}
    \label{fig:example}
\end{figure}
\begin{figure}[!h]
    \begin{subfigure}[t]{.45\linewidth}
        \includegraphics[width=\linewidth]{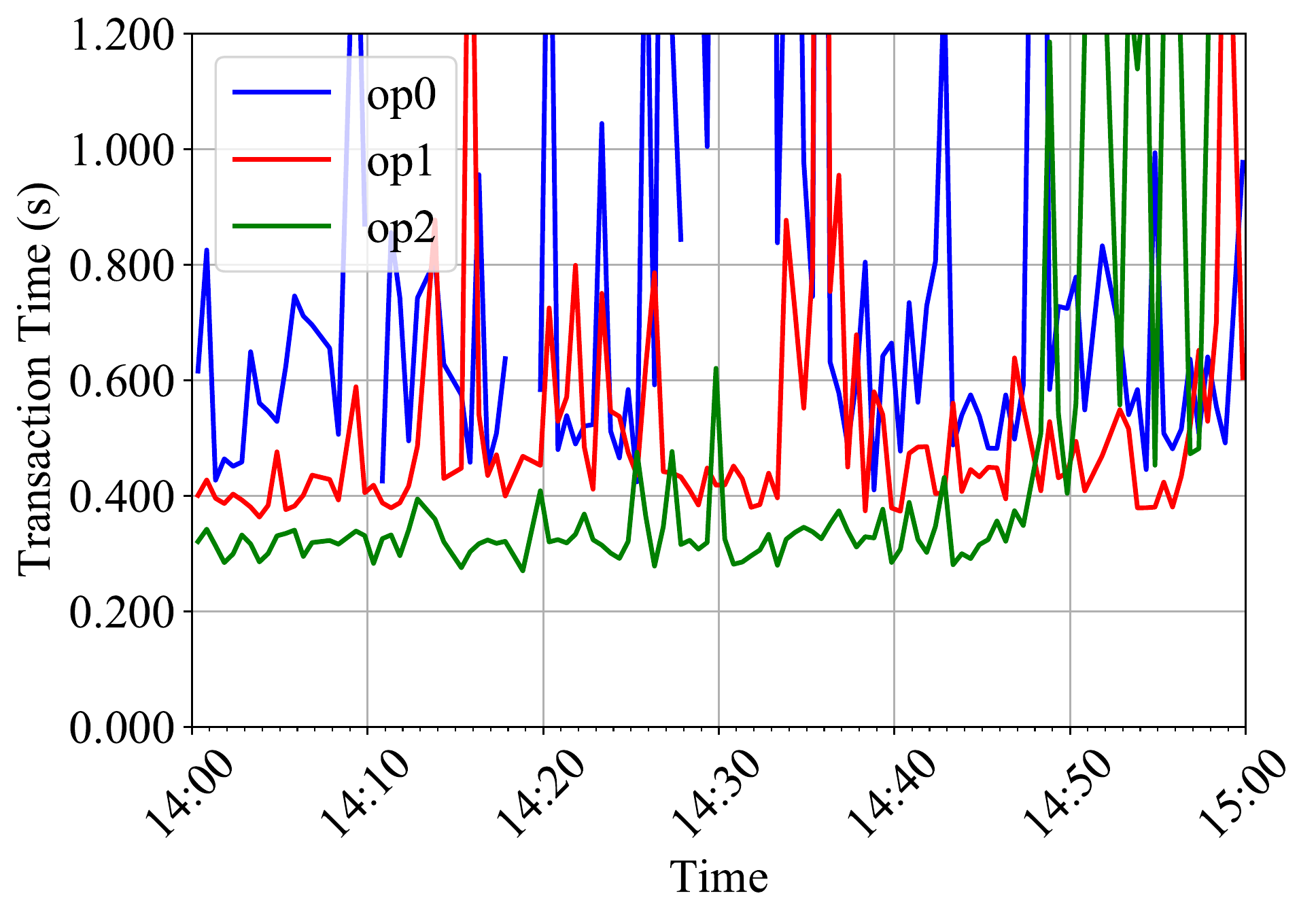}
    \end{subfigure}
    \begin{subfigure}[t]{.45\linewidth}
        \includegraphics[width=\linewidth]{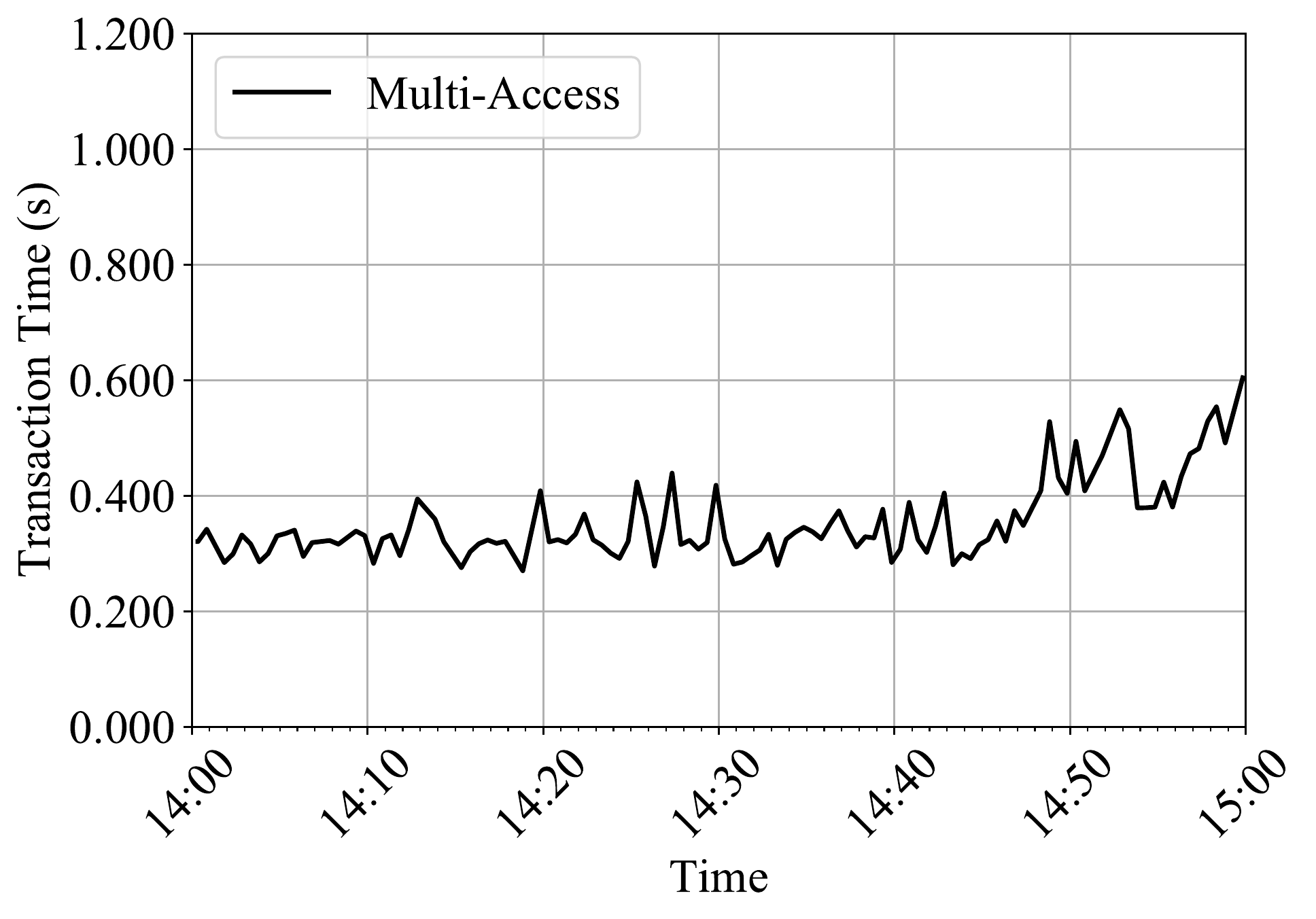}
    \end{subfigure}
    \caption{Typical example of an experiment run with HTTPS.}
    \label{fig:example}
\end{figure}

\begin{figure*}[!h]
    \begin{subfigure}[t]{0.3\textwidth}
        \includegraphics[width=\textwidth]{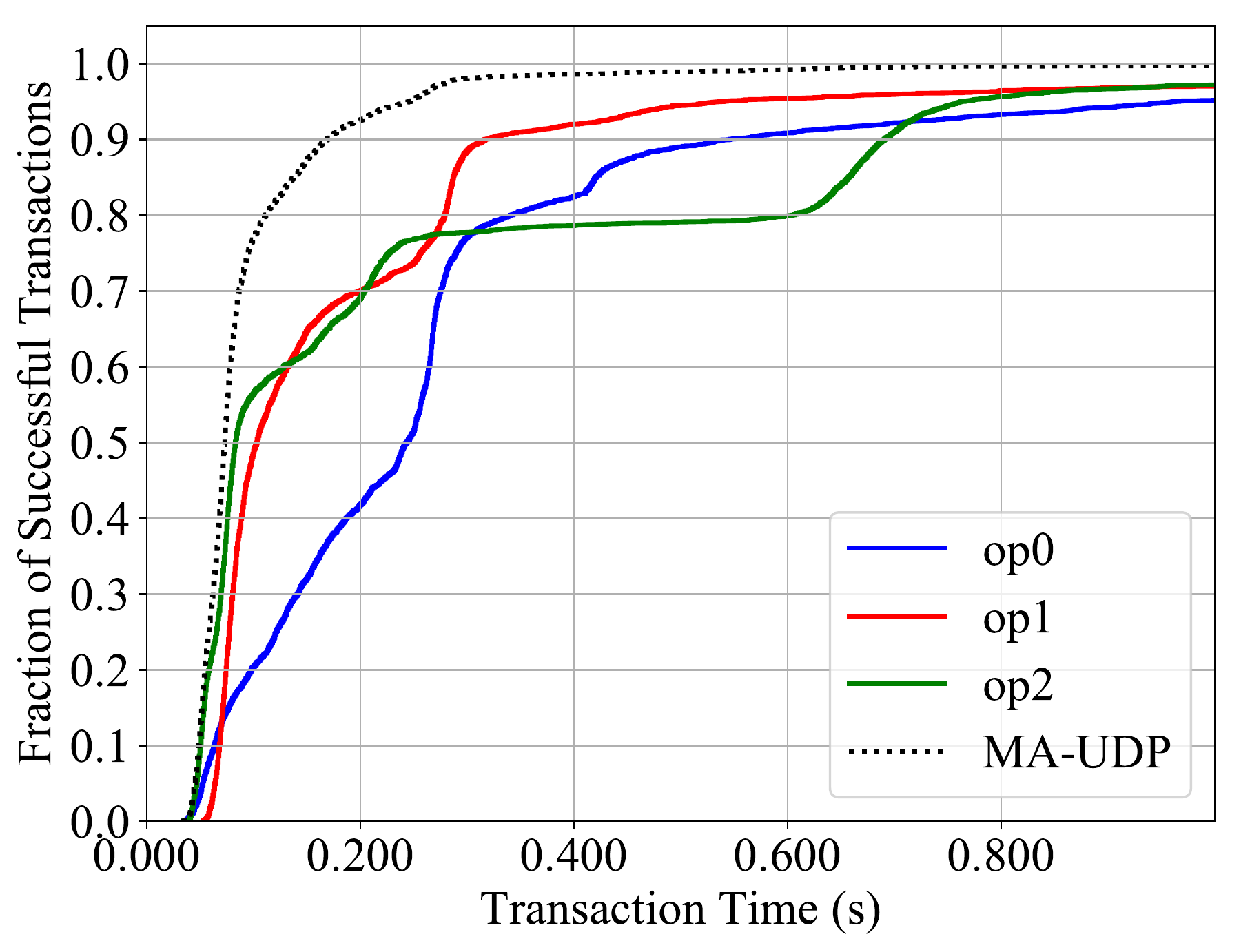}
        \caption{UDP}
    \end{subfigure}
    \begin{subfigure}[t]{0.3\textwidth}
        \includegraphics[width=\textwidth]{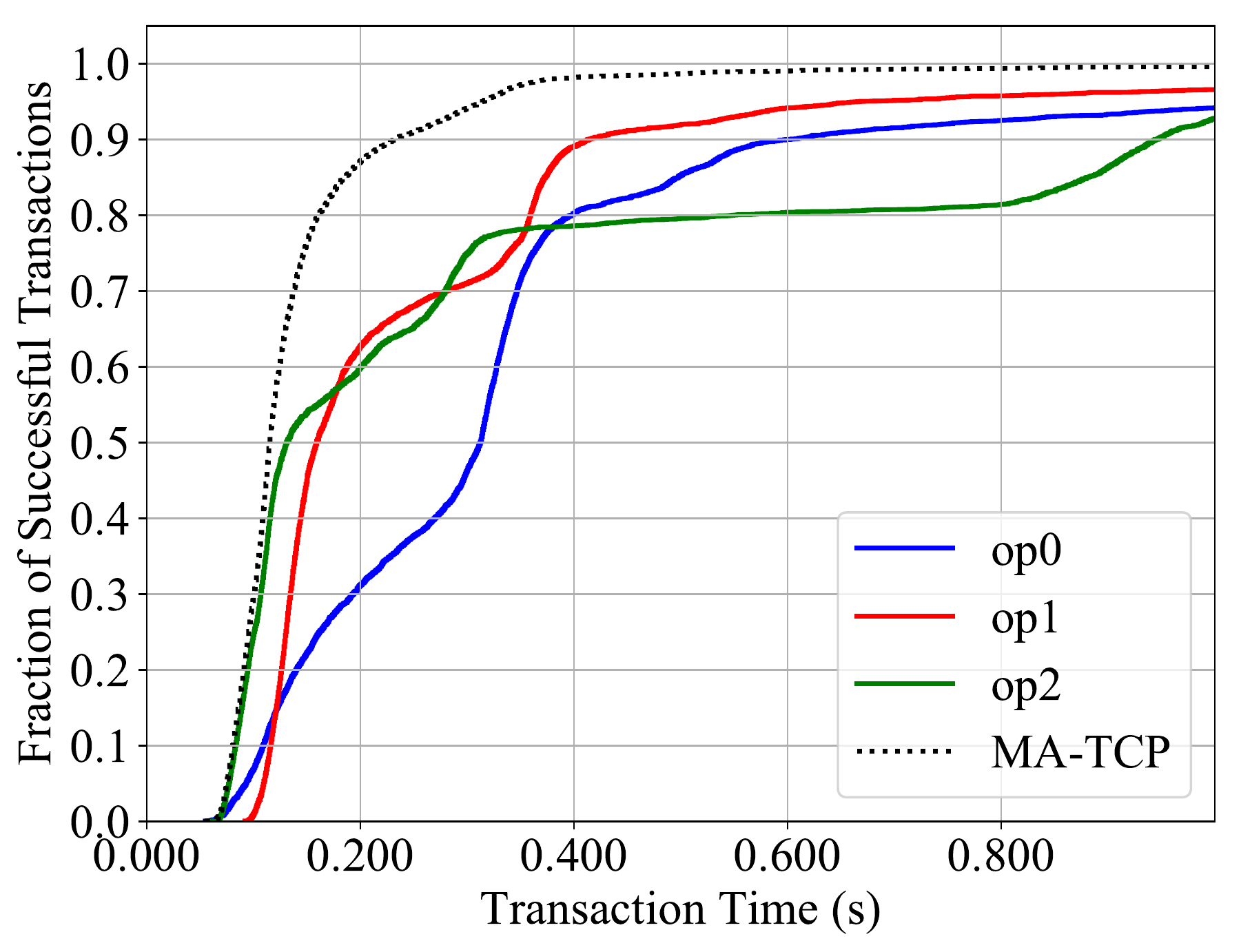}
        \caption{TCP}
           \label{fig:peropTCP}
    \end{subfigure}
    \begin{subfigure}[t]{0.3\textwidth}
        \includegraphics[width=\textwidth]{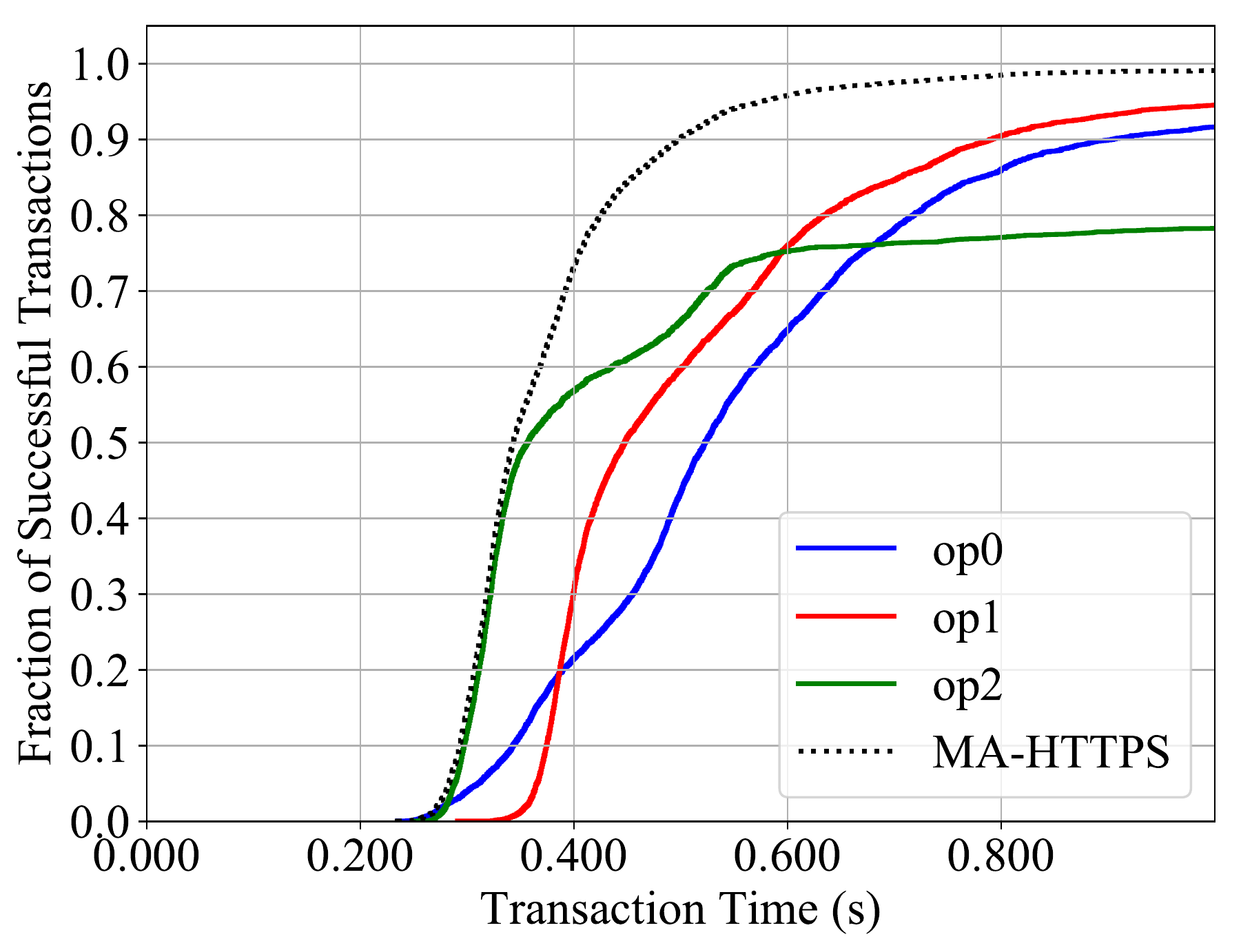}
      \caption{HTTPS}
    \end{subfigure}

    \caption{ECDF of transaction times for each operator and protocol.}
    \label{fig:perop}
\end{figure*}

We are mainly interested in transactions that were performed at the same time on different operators. Despite our efforts to syncronise the transactions, we observe that sometimes the time difference between the start of the transactions performed in a same round can be as high as $300$~ms. This is explained by the design of the MONROE nodes, where two interfaces are on one separate machine (\emph{head} node), and the third interface is on another separate machine (\emph{tail} node). Both machines are seen as one node in our experiment, and are co-located in a same box in the buses. The synchronisation between the head node and the tail node was not always precise and we decided to only consider transactions that started within $10$~ms from each other. All results in this section are obtained from the filtered dataset containing only synchronised transactions.

\subsection{Example}

Fig.~\ref{fig:example} shows an example of experiment run, illustrating the variation of the transaction time when using UDP with different operators simultaneously. The gaps in the curves occur when the transaction fails, because of lost packets, lack of network signal, or due to a transaction time-out. Intuitively, using all operators simultaneously (Multi-Access) could potentially reduce the transaction time, and Fig.~\ref{fig:example} confirms that doing so indeed keeps the transaction time shorter, while also reducing the number of transaction failures. We measure the performance of multi-access by picking the best transaction time value in each round. 

\begin{figure}[h]
\centering
\includegraphics[width=0.45\hsize]{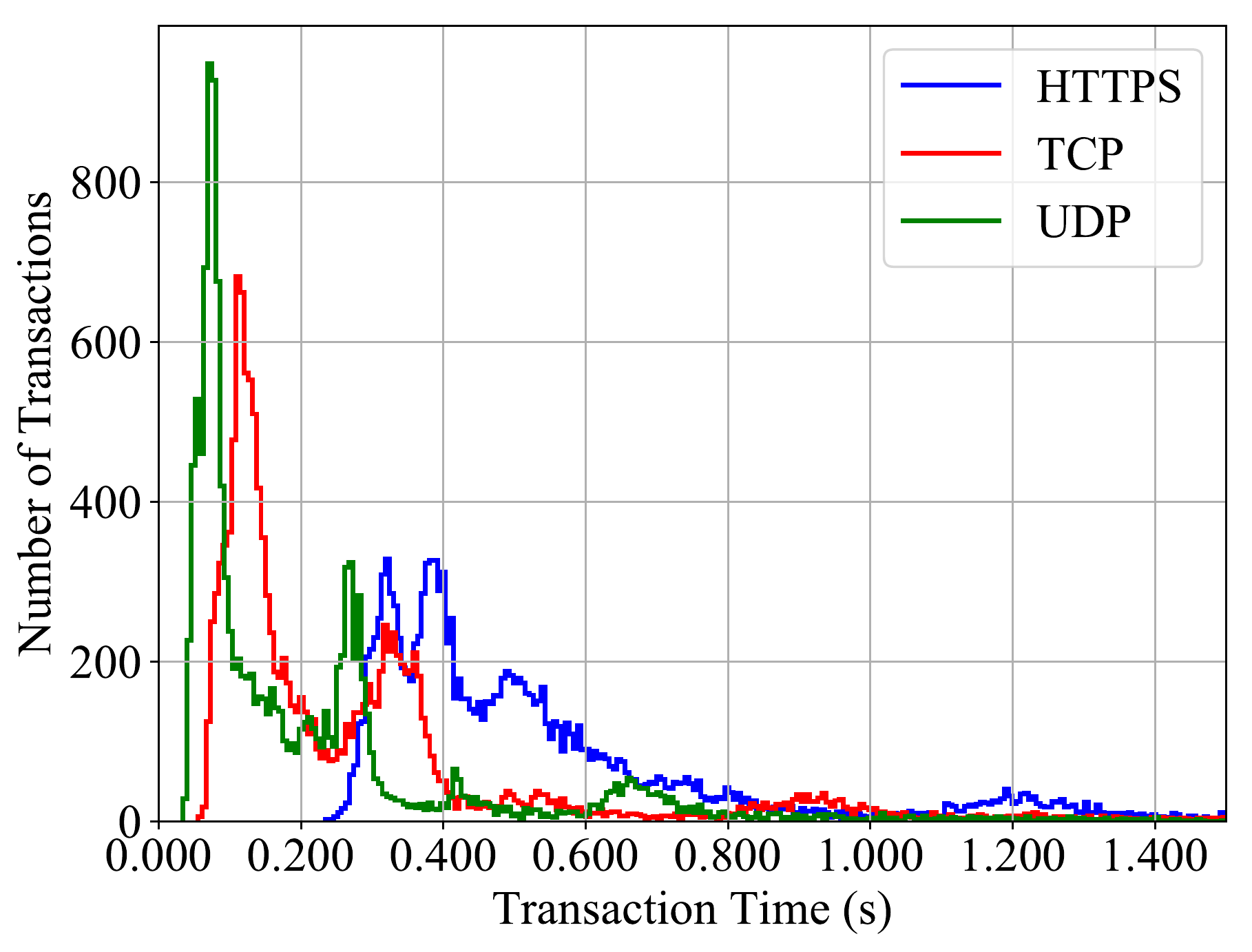}
\caption{Distribution of the transaction times for each protocol.}
\label{fig:transactionTimeDist}
\end{figure}

\begin{table*}[h]
\renewcommand{\arraystretch}{1.3}
\caption{Transaction time for each protocol.}
  \label{tab:transactiontime}
 \centering
    \begin{tabular}{ | r | c | c | c | c | c | c | c | c |}
  \hline			
        \bfseries Protocol 
        & \multicolumn{1}{p{1.5cm}|}{\bfseries Attempted transactions}
        & \multicolumn{1}{p{1.5cm}|}{\bfseries Successful transactions} 
        & \bfseries Mean 
        & \multicolumn{1}{p{1.5cm}|}{\bfseries Standard deviation}
        & \bfseries Median 
        & \bfseries Mininum
        & \multicolumn{1}{p{1.4cm}|}{\centerline{\bfseries 90\% quantile}}
        & \bfseries Maximum\\
  \hline			
  \hline
\bfseries UDP & 14448 & 13770 & 0.273 & 0.465 & 0.135 & 0.034 & 0.643 & 5.871\\
\rowcolor{Gray}
\bfseries MA-UDP & 14448 & 14364 & 0.103 & 0.180 & 0.073 & 0.034 & 0.169 & 5.242\\
\bfseries TCP & 15009 & 14765 & 0.350 & 0.512 & 0.193 & 0.055 & 0.813 & 5.983\\
\rowcolor{Gray}
\bfseries MA-TCP & 15009 & 14988 & 0.148 & 0.175 & 0.115 & 0.055 & 0.232 & 5.607\\
\bfseries HTTPS & 15387 & 15102 & 0.612 & 0.504 & 0.469 & 0.234 & 1.145 & 5.845\\
\rowcolor{Gray}
\bfseries MA-HTTPS & 15387 & 15366 & 0.385 & 0.179 & 0.343 & 0.234 & 0.499 & 5.519\\
  \hline  
\end{tabular}
\end{table*}

\subsection{Transaction time}

To quantify the potential gain of using multi-access, we first aggregate all transactions in the dataset and characterise the transaction time values for each protocol. Fig.~\ref{fig:transactionTimeDist} shows the distribution of the transaction times in the dataset, to give an overall insight on what can be expected from today's cellular networks on the road. We clearly observe a bimodal distribution for all protocols, explained by the concurrent usage of both LTE and 3G. Table~\ref{tab:transactiontime} provides statistics on the transaction times measured for each protocol, and shows that the median value for UDP is $135$~ms, while $193$~ms and $469$~ms can be expected from TCP and HTTPS, respectively. That UDP outperforms TCP and HTTPS in terms of transaction time was expected, as both less packets and less round trips are required for the transaction to complete. Table~\ref{tab:transactiontime} also shows that using multi-access reduces the standard deviation of the transaction times, hence guaranteeing a more stable connection. The maximum value is limited by the $6$~second time-out. The distribution and characteristics of the transaction times as shown in Fig.~\ref{fig:transactionTimeDist} and Table~\ref{tab:transactiontime} provide an insight on what time-critical applications can expect from the network.

\begin{figure}[h]
\centering
\includegraphics[width=0.45\hsize]{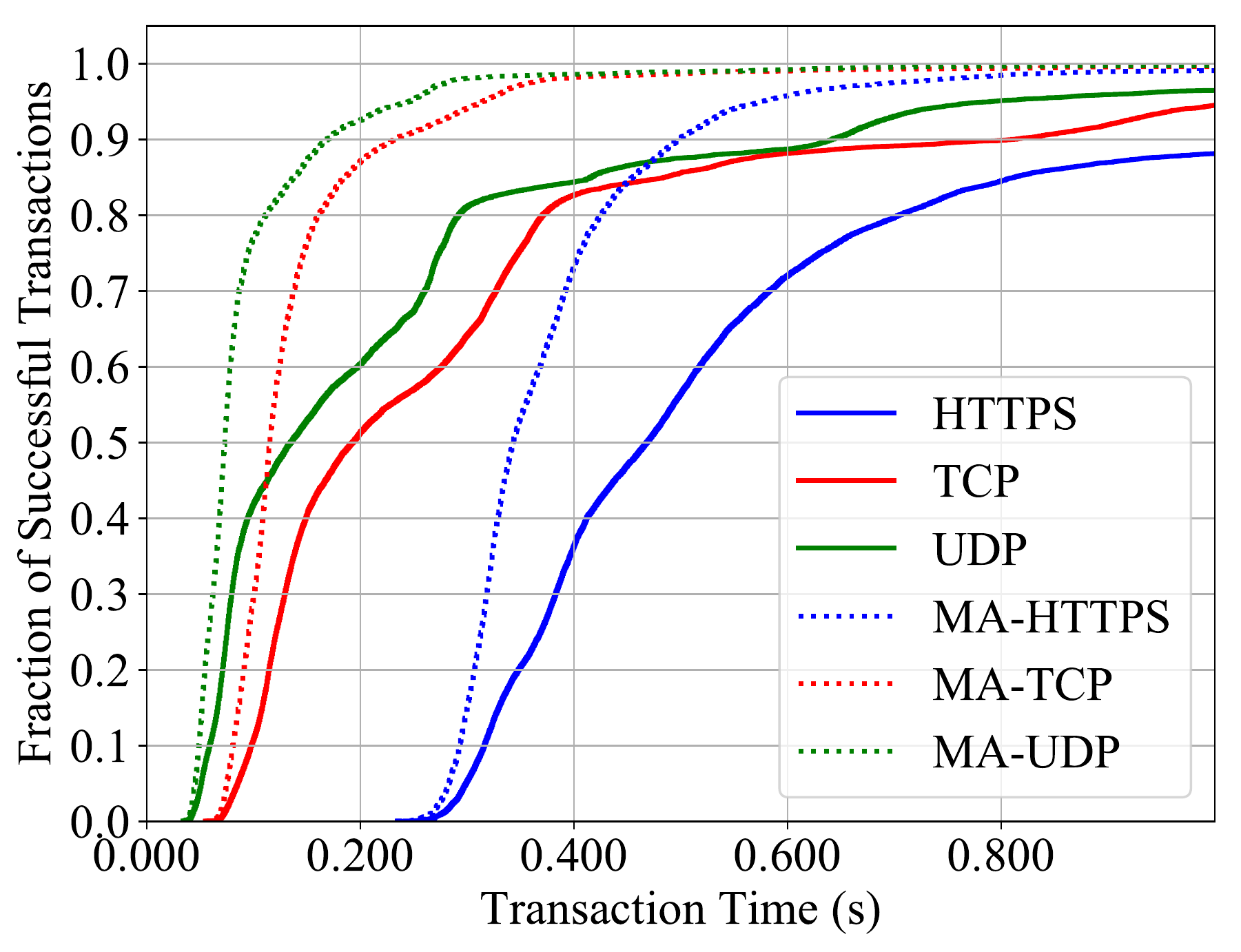}
\caption{ECDF of the transaction times for each protocol.}
\label{fig:transactionTime}
\end{figure}

Fig.~\ref{fig:transactionTime} shows an empirical cumulative distribution function (ECDF) of the transaction times for each protocol, confirming our intuition that using multi-access reduces the transaction times, for all protocols tested. For instance, we observe that nearly $80$ \% of UDP transactions were successfully completed in less than $100$~ms, while $80$ \% of UDP transactions without multi-access take up to $300$~ms to complete. Table~\ref{tab:transactiontime} also shows that the median value for transaction times is almost halved when using multi-access with UDP, from $135$~ms to $73$~ms. The standard deviation values are also much lower for all protocols tested when using multi-access, offering a better stability for the transaction times.

\subsection{Availability}

Not only does multi-access allows shorter transaction times, it also significantly increases the success rates of the transactions for all protocol tested. Table~\ref{tab:successrate} shows the success rates of the transactions for different time limits. While the client time-out value has been set to $6$~seconds, as explained in Section~\ref{subsec:exp}, we can simulate shorter time limits in post-processing, such as $1$~second and $200$~ms, by considering longer transactions as failed transactions. Table~\ref{tab:successrate} shows that multi-access increases the success rate by an additional $4.1$\% for UDP. For time-critical applications that require a very short time limit of $200$~ms, the success rate of the transactions when using multi-access bumps from $57.4$\% to $92.0$\% for UDP, and from $50.4$\% to $86.9$\% for TCP. Note that HTTPS is not feasible for such time limits.

\begin{table}[h]
\renewcommand{\arraystretch}{1.3}
\caption{Availability for each protocol and time limit.}
  \label{tab:successrate}
 \centering
\begin{tabular}{ | r | c | g | c | g | c | g |}
  \hline			
     \multicolumn{1}{|p{0.6cm}|}{\bfseries Time limit} & \bfseries UDP & \multicolumn{1}{p{0.6cm}|}{\bfseries MA-UDP} & \bfseries TCP & \multicolumn{1}{p{0.6cm}|}{\bfseries MA-TCP} & \bfseries HTTPS & \multicolumn{1}{p{0.8cm}|}{\bfseries MA-HTTPS}\\
  \hline			
  \hline
\bfseries 6~s & 95.3\% & 99.4\% & 98.4\% & 99.9\% & 98.1\% & 99.9\% \\
\bfseries 1~s & 91.9\% & 99.1\% & 93.0\% & 99.4\% & 86.5\% & 98.9\% \\
\bfseries 0.2~s & 57.4\% & 92.0\% & 50.4\% & 86.9\% & 0.0\% & 0.0\% \\
  \hline  
\end{tabular}
\end{table}

\subsection{Comparing multi-access with each operator}
The potential gain of using multi-access can also be compared to using each operator individually. Fig.~\ref{fig:perop} shows the distribution of transaction times for each protocol tested, and for each operator. The figure shows that using multi-access yields significantly shorter transaction times than with any of the operators. For instance, Fig.~\ref{fig:peropTCP} shows that when using multi-access with TCP, almost $90$\% of transactions are completed in less than $200$~ms, whereas the transactions completed under $200$~ms for the best operator only amount for around $63$\% of the transactions.    

\subsection{Transactions with larger messages}

\begin{figure}[h]
\centering
\includegraphics[width=0.6\hsize]{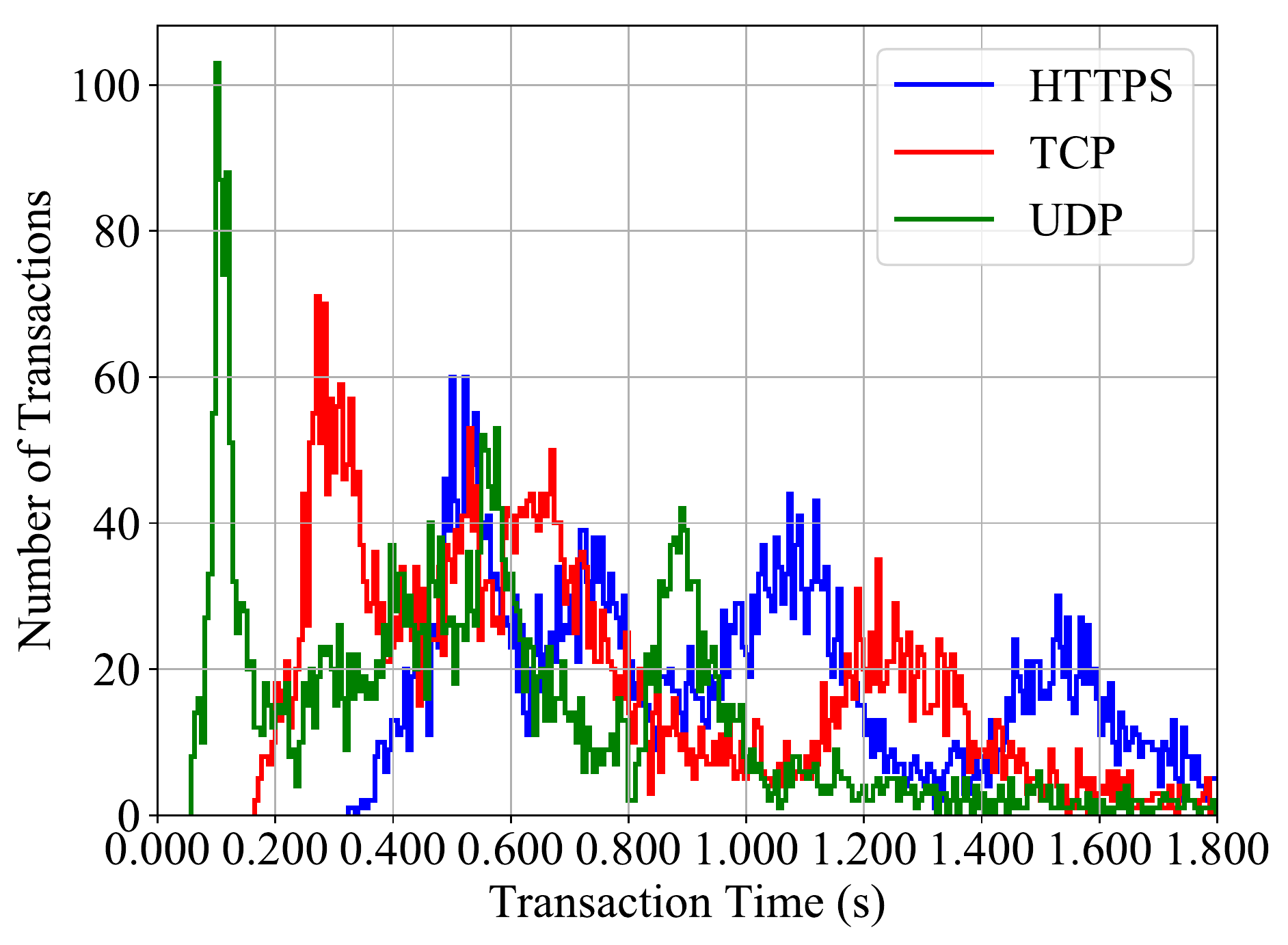}
\caption{Distribution of the transaction times for each 50 KB transaction.}
\label{fig:transactionTime50Dist}
\end{figure}

\begin{figure}[h]
\centering
\includegraphics[width=0.6\hsize]{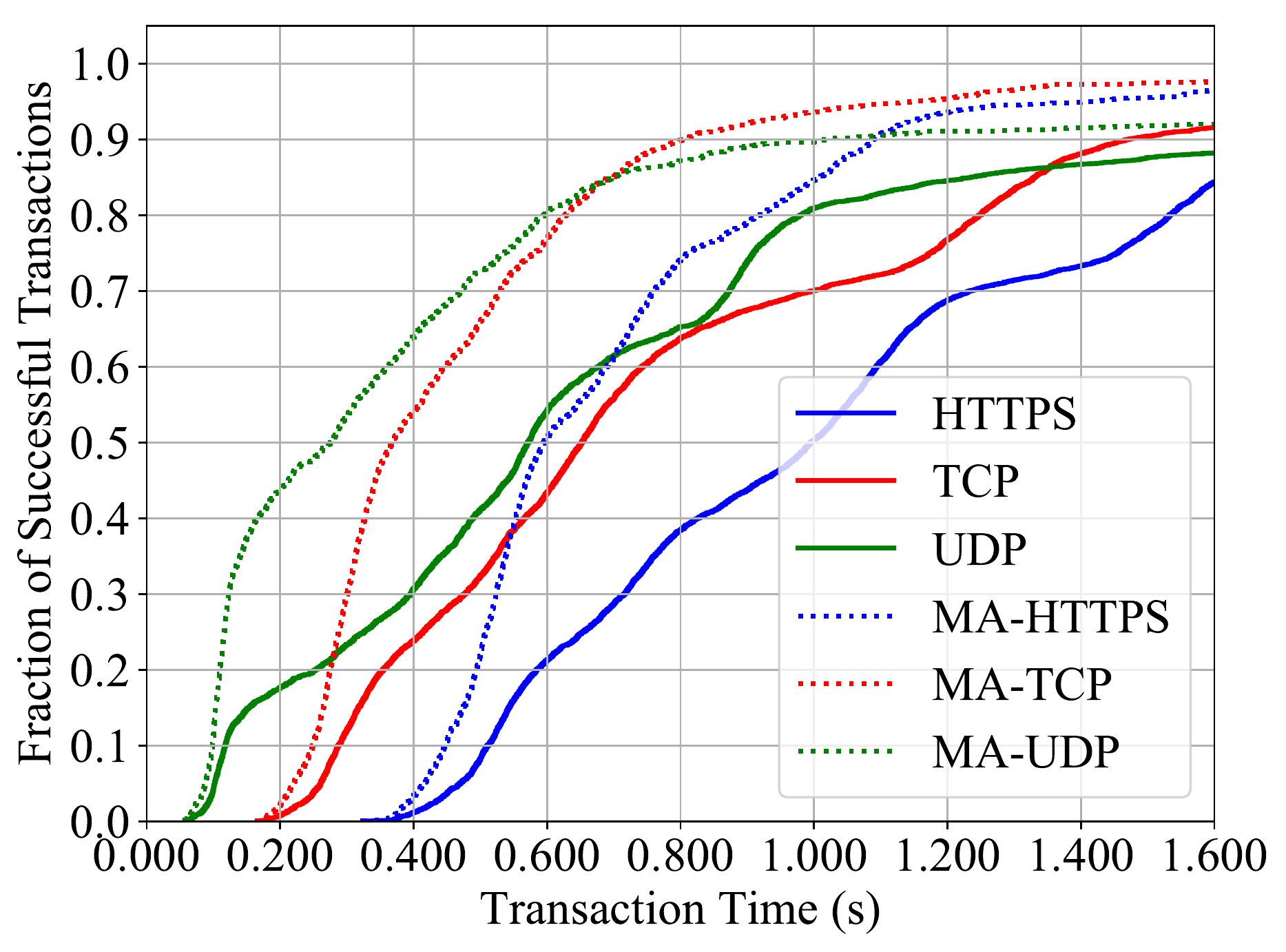}
\caption{ECDF of the transaction times for each 50 KB transaction.}
\label{fig:transactionTime50}
\end{figure}

As more and more sensors are embedded in modern cars, larger amount of data could be uploaded in the future, possibly for time-critical \cits\ applications. To assess how the cellular network could handle such large transactions, we ran similar experiments with a $50$~KB message. Depending on the protocol, almost $6,000$ synchronised transactions (under $10$~ms from each other) where measured, as shown in Table~\ref{tab:transactiontime50}. The table also shows that, as expected, the transaction time measured is longer, with a median value of $572$~ms for UDP, against $135$~ms for $5.6$~KB messages. We observe a multi-modal distribution of transaction times that is also more spread, and that shows three to four peaks depending on the protocol, as shown in Fig.~\ref{fig:transactionTime50Dist}. However, Fig.~\ref{fig:transactionTime50} shows that using multi-access is still significantly beneficial to all protocols tested. For instance, while almost $90$\% of the successful transactions are completed under $800$~ms when using multi-access for both UDP and TCP, only around $65$\% of the successful transactions are completed under $800$~ms without multi-access. 

\begin{table*}[h]
\renewcommand{\arraystretch}{1.3}
\caption{Transaction time when uploading 50 KB messages, for each protocol.}
  \label{tab:transactiontime50}
 \centering
    \begin{tabular}{ | r | c | c | c | c | c | c | c | c |}
  \hline			
        \bfseries Protocol 
        & \multicolumn{1}{p{1.5cm}|}{\bfseries Attempted transactions}
        & \multicolumn{1}{p{1.5cm}|}{\bfseries Successful transactions} 
        & \bfseries Mean 
        & \multicolumn{1}{p{1.5cm}|}{\bfseries Standard deviation}
        & \bfseries Median 
        & \bfseries Mininum
        & \multicolumn{1}{p{1.4cm}|}{\centerline{\bfseries 90\% quantile}}
        & \bfseries Maximum\\
  \hline			
  \hline
\bfseries UDP & 5925 & 4561 & 0.991 & 1.368 & 0.572 & 0.057 & 1.973 & 5.998\\
\rowcolor{Gray}
\bfseries MA-UDP & 5925 & 4920 & 0.691 & 1.358 & 0.277 & 0.057 & 1.016 & 5.993\\
\bfseries TCP & 5976 & 5577 & 0.870 & 0.737 & 0.652 & 0.165 & 1.471 & 5.938\\
\rowcolor{Gray}
\bfseries MA-TCP & 5976 & 5892 & 0.511 & 0.480 & 0.368 & 0.165 & 0.804 & 5.759\\
\bfseries HTTPS & 5778 & 5470 & 1.139 & 0.737 & 0.996 & 0.324 & 1.796 & 5.945\\
\rowcolor{Gray}
\bfseries MA-HTTPS & 5778 & 5718 & 0.745 & 0.478 & 0.595 & 0.324 & 1.089 & 5.807\\
  \hline  
\end{tabular}
\end{table*}

We observe from Table~\ref{tab:successrate50} that for such large messages, the success rate drops dramatically for TCP and HTTPS when the time limit is one second: from $93$\% to $65.3$\% for TCP, and from $86.5$\% to $47.6$\% for HTTPS. However, using multi-access allows to maintain a high success rate under the same conditions, with $92.3$\% for UDP and $83.7$\% for HTTPS.

\begin{table}[h]
\renewcommand{\arraystretch}{1.3}
\caption{Availability with 50 KB messages \\for each protocol and time limit.}
  \label{tab:successrate50}
 \centering
\begin{tabular}{ | r | c | g | c | g | c | g |}
  \hline			
     \multicolumn{1}{|p{0.6cm}|}{\bfseries Time limit} & \bfseries UDP & \multicolumn{1}{p{0.6cm}|}{\bfseries MA-UDP} & \bfseries TCP & \multicolumn{1}{p{0.6cm}|}{\bfseries MA-TCP} & \bfseries HTTPS & \multicolumn{1}{p{0.8cm}|}{\bfseries MA-HTTPS}\\
  \hline			
  \hline
\bfseries 6~s & 77.0\% & 83.0\% & 93.3\% & 98.6\% & 94.7\% & 99.0\% \\
\bfseries 1~s & 62.2\% & 74.4\% & 65.3\% & 92.3\% & 47.6\% & 83.7\% \\
\bfseries 0.2~s & 13.6\% & 36.3\% & 0.7\% & 2.1\% & 0.0\% & 0.0\% \\
  \hline  
\end{tabular}
\end{table}

\section{Discussion and Ongoing Work}
\label{sec:discussion}
Connected cars can make road traffic safer and more efficient, but require the mobile networks to handle time-critical applications. Some \cits\ applications require transaction times below $100$~ms~\cite{Karagiannis:cst11}. Multi-access reduces the median transaction time for UDP by $46$\%, from $135$~ms down to $73$~ms, well below the $100$~ms limit. With multi-access, nearly $80$\% of UDP transactions complete under $100$~ms.

Transactions completed over longer times can still be valuable to some other \cits\ applications. For instance, when a vehicle sends a warning message about a wild animal on the road, other vehicles that will reach the same location in a minute can receive the warning after several seconds. Here network availability is crucial, and delivering the message is important even when delayed. Our measurements show that multi-access can significantly improve availability to $99.9$\%, from $98.4$\% for TCP and from $98.1$\% for HTTPS. 

\begin{figure}[h]
\centering
\includegraphics[width=0.6\hsize]{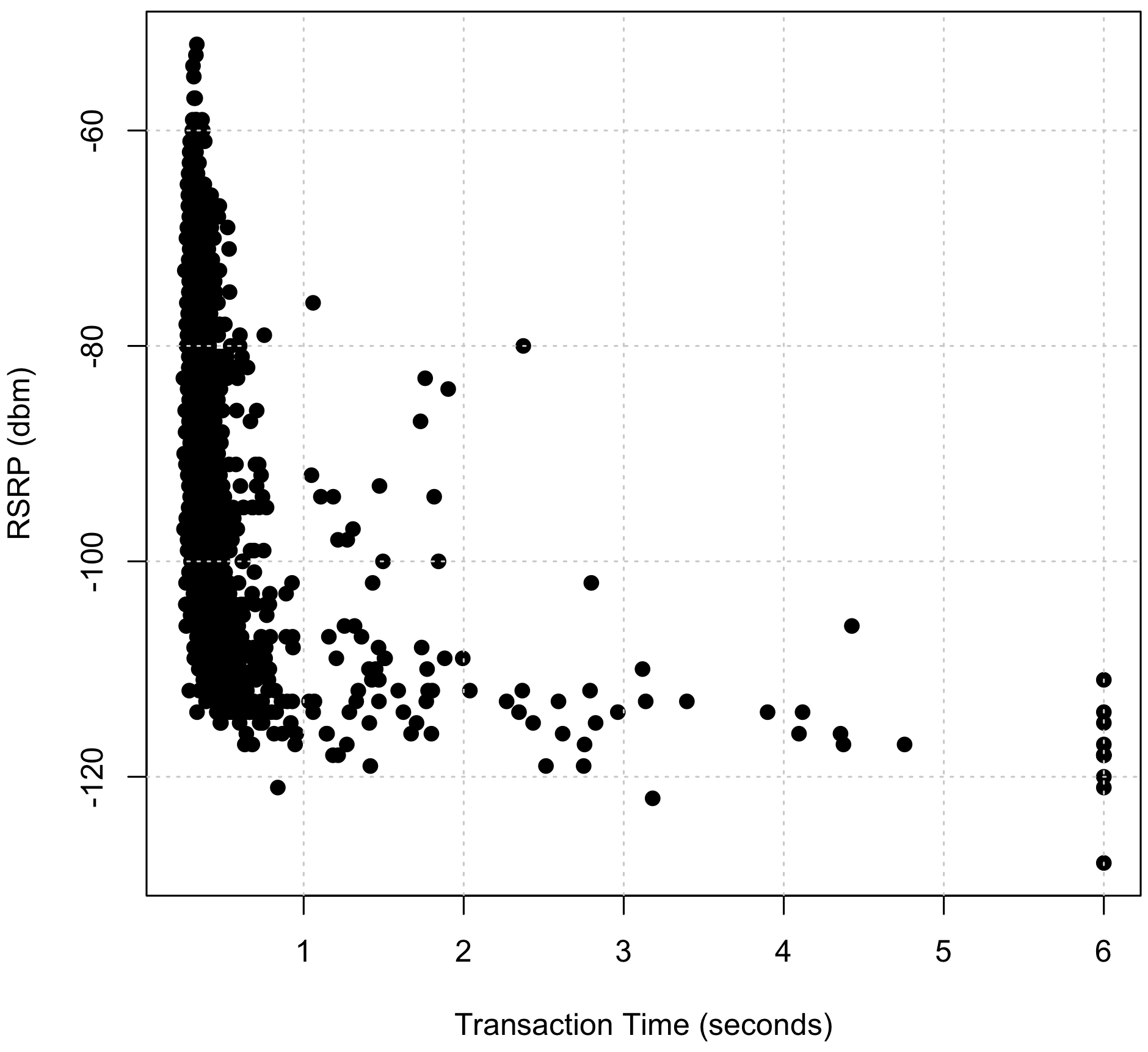}
\caption{Reference Signal Received Power (RSRP) plotted against transaction time for HTTPS transactions over LTE. The time for failed transactions have been set to six seconds. The Spearman's rank correlation coefficient is -0.48.}
\label{fig:scatterplotRSRP}
\end{figure}

Multi-access inflicts an expensive overhead on the cellular network, by duplicating data traffic over several operators. While the potential gain demonstrated by our study might in some cases be worth the price, we are currently investigating ways to avoid this overhead by selecting the best network operator when performing a transaction. The dataset we collected includes a rich set of meta-data, such as the signal strength and the GPS coordinates before each transaction. Finding correlations between this meta-data and the transaction time would help selecting the best network for the transaction. It would also help explaining the performance difference between operators, beyond the disparity in radio technology deployment. Fig.~\ref{fig:scatterplotRSRP} shows a promising correlation between the Reference Signal Received Power (RSRP) and the transaction time for HTTPS.  

The potential gain of multi-access measured in this study can be seen as an upper limit for improvement. Our ongoing investigation aims at getting as close as possible to that upper limit, with an in-depth analysis of the correlation between the transaction times and all other features. Studying this correlation will help better predict the network performance of each operator, and pick the best one for the transaction.

\section{Conclusions}
\label{sec:conclusions}
Many \cits\ applications will have to rely on cellular networks. Depending on the exact time and reliability constraints, we outlined the conditions under which time-critical applications can be handled by existing networks. The characterisation of transaction times and network availability under those conditions helps understanding what can be expected from existing networks on the road. The median transaction time that can be expected from the network when sending a typical warning message is $135$~ms over UDP, $193$~ms over TCP, and $469$~ms over HTTPS. We demonstrated the significant potential gain of multi-access, reducing those values down to $73$~ms, $115$~ms, and $343$~ms, respectively. We also showed that multi-access increases availability from $57.4$\% to $92.0$\% for time-critical applications when using UDP, and from $50.4$\% to $86.9$\% when using TCP. Those promising results are paving the way to our ongoing investigation on how to select the best network for each transaction, hence optimising the performance of \cits\ services on existing networks.

\section*{Acknowledgments}
This work is funded by the European Union's Horizon 2020 research and innovation programme under grant agreement No. 644399 (MONROE) through the open call project FELICIA. The views expressed are solely those of the authors.


\bibliographystyle{IEEEtran}
\bibliography{connectedvehicles.bib}

\end{document}